\documentclass[preprint,eqsecnum,aps,showpacs]{revtex4}
\usepackage{dcolumn}
\usepackage{graphicx}
\usepackage{epsfig}
\usepackage{float}
\usepackage{latexsym}
\usepackage{rotating}
\begin{document}
%\addtolength{\topmargin}{25pt}
\title{Kinematics of deformable media}
\author{
Anirvan Dasgupta\footnote{Electronic address: {\em anir@mech.iitkgp.ernet.in}}${}^{}$}
\affiliation{Department of Mechanical Engineering and Centre for Theoretical Studies  \\
Indian Institute of Technology, Kharagpur 721 302, India}
\author{
Hemwati Nandan\footnote{Electronic address: {\em hnandan@cts.iitkgp.ernet.in}}${}^{}$}
\affiliation{Centre for Theoretical Studies  \\
Indian Institute of Technology, Kharagpur 721 302, India}
\author{
Sayan Kar\footnote{Electronic address: {\em sayan@cts.iitkgp.ernet.in}}${}^{}$}
\affiliation{Department of Physics and Centre for Theoretical Studies  \\
Indian Institute of Technology, Kharagpur 721 302, India}
\begin{abstract}
We investigate the kinematics of deformations in two and
three dimensional media by explicitly solving (analytically) the evolution equations (Raychaudhuri equations) for the expansion, shear and rotation associated with the deformations. The analytical solutions allow us to study the dependence of the kinematical  quantities on initial conditions. In particular, we are able to identify regions of the space of initial conditions that lead to a singularity in finite time.  Some generic features of the deformations are also discussed in detail. We conclude by indicating the feasibility and utility of a similar
exercise for fluid and geodesic flows in flat and curved spacetimes.
\end{abstract}

\maketitle

\section{Introduction}
Deformations (elastic or otherwise) under applied external stress is a 
well--studied subject \cite{ll}. It is known that the kinematics of such
deformations can be understood by analysing the behaviour of 
the kinematical quantities, namely the {\em isotropic expansion}, the 
{\em shear} and the {\em rotation or vorticity}. 
However, as far as we know, most of these investigations 
do not explicitly address the question of evolution of
the expansion, rotation or shear in {\em time} by solving the 
associated initial value problem for the system of differential
equations involving these variables. 
We believe that such a study could shed light on the
role of initial conditions (more precisely, the initial values of
the expansion, rotation and shear at some time) 
in the kinematics of deformations.
Additionally, many subtleties (integrals of motion, critical values, 
appearance of singularities etc.) in the nature of the evolution of
these quantities may emerge if one is able to obtain exact
solutions or even scan the full domain of numerical solutions
of the relevant coupled, nonlinear system of evolution equations
involving expansion, shear and rotation.

The equations we will discuss in this article are, essentially,
the Raychaudhuri equations \cite{review}, so prominent and
well-known in the study of spacetime singularities in gravitation
and cosmology \cite{hawk}-\cite{joshi}. In the context of
gravitation, the central result that emerges from the Raychaudhuri
equations, is the notion of geodesic focusing. The analysis
which is based on an inequality (in the first order approach) 
or the theory of second order ordinary differential
equations (in the second order approach) proves the seemingly
simple statement: {\em if gravity is attractive then geodesics
must focus to a singularity in the congruence (which, may or may not
be a genuine spacetime singularity)}. In the broader context of
Riemannian geometry, focusing in a geodesic flow arises provided
one obeys the appropriate convergence condition.  

In this article, we, however,
will restrict ourselves to a background flat space (not spacetime)
and first deal with two spatial dimensions, and then with three
dimensions. The family of worldlines (the so-called geodesic
congruence) will be replaced, in our case (i.e. for deformable
media), by the deformation vectors, with the
time derivative of the deformation vector replacing the {\em
tangent vector} to the {\em geodesic flow}. The {\em geodesic
equation} used to generate the geodesic flow, which is
normally used as an input in deriving the Raychaudhuri equations
(one can also write equations for non--geodesic flows too, but we
do not consider them here) will now be replaced by
a second order (in time) equation for the deformation vector. In
other words, we shall exactly follow the scheme of derivation of
the Raychaudhuri equations in the context of General Relativity,
keeping in mind the abovementioned parallels and restrictions. We
mention here that by virtue of being a geometric statement, the
Raychaudhuri equations are applicable to a myriad of diverse
situations ranging from deformations in media, fluid flow,
geodesic congruences etc. Our aim, as mentioned before, is to
analyse, in all possible detail, the kinematics of deformations
of a medium without and with stiffness and viscosity. In particular,
one of the features we investigate is: how does the appearance of a
singularity depend on the initial conditions on the expansion, shear or
rotation.
The paper is broadly divided into two parts. In Section II we talk about
two dimensional media, while in Section III we focus attention on
three dimensional media. In each part, we write down the relevant equations
involving the expansion, rotation and shear and find analytic
solutions for these variables. 
Subsequently, we try to understand the solutions obtained by
looking for generic features and their dependence on initial conditions.

\section{Deformations in two dimensional media}

\subsection{Expansion, rotation and shear}

Consider, following {\cite{toolkit}}, a two dimensional deformable medium. 
An explicit example could be a thin sheet of rubber or plasticine or any
putty--like modeling material. 
Imagine that, at $t=0$, an initial `velocity' field (time rate of change
of deformation) is specified for the 
medium. This will correspond to an expansion, a shear and a rotation (ESR)
of the medium. Alternatively, one may specify
(in place of initial velocity field) initial conditions on the ESR 
themselves. It may be recalled that the study of the evolution of the 
ESR variables of
a medium is important in many areas (such as General Relativity). To 
proceed further,
one needs to derive the evolution equations of the deformation in terms of the
ESR variables. This is discussed below.
Let us denote the deformation of the medium in terms of the deformation vector 
$\xi^{i}$ (where $i=1,2$),
which represents the vector joining any two infinitesimally separated points 
of the medium.
The time rate of change of $\xi^{i}$ may be expressed, for small time intervals, as
(see Appendix A)
\begin{equation}
\frac{d\xi^{i}}{dt} = B^i_{\,\,j} (t) \xi^{j}+{\cal O}(\Delta t^2),
\label{devveq}
\end{equation}
where $B^{i}_{\,\,j}(t)$ is an arbitrary second rank tensor
characterising the time evolution of the deformation vector. 
Differentiating (\ref{devveq}) with respect to time and using (\ref{devveq}) to eliminate $\dot{\xi}^i$,
we obtain
\begin{equation}
\left (\frac{dB^i_{\,\,j}}{dt} + B^i_{\,\,k}B^k_{\,\,j} \right ) \xi^j={\ddot\xi}^i.
\label{ddotxi}
\end{equation}
Before we write down the evolution equations for the ESR, let us first define 
mathematically the quantities of interest.
The arbitrary second rank tensor $B^i_{\,\,j}$ can be decomposed into
its trace, symmetric traceless and antisymmetric parts which will
correspond to the isotropic expansion (trace scalar),
shear (symmetric traceless tensor) and rotation (antisymmetric tensor). 
This decomposition is given as:
\begin{equation}
B_{ij} = \frac{1}{2}{\theta} \, \delta_{ij} +\sigma_{ij} + \omega_{ij}, \label{bij}
\end{equation}
where $\theta$, $\sigma_{ij}$ and $\omega_{ij}$ represent, respectively, the 
expansion, shear and rotation. We can explicitly write (\ref{bij}) in terms of the
following $2\times 2$ matrices:
\begin{equation}
\frac{1}{2}\theta\delta_{ij}=
\left(
\begin{array}{cc}
\frac{1}{2}\theta & 0\\
0&  \frac{1}{2}\theta\\
\end{array} \right),~~~~~~
\sigma_{ij}=
\left(
\begin{array}{cc}
\sigma_{+} & \sigma_{\times} \\
\sigma_{\times} & -\sigma_{+} \\
\end{array} \right), ~~~~~~~
\omega_{ij}=
\left(
\begin{array}{cc}
0 & -\omega \\
\omega& 0 \\
\end{array} \right),
 \label{ctrans}
\end{equation}
where we have denoted the shear components as $\sigma_+$ and $\sigma_\times$,
and the only rotation component as $\omega$. Thus, these four quantities
$\theta$, $\sigma_+$, $\sigma_\times$ and $\omega$ characterise any
deformation of the two dimensional deformable medium as a function 
of time. One can find
useful discussions on the geometrical interpretation of each of these 
quantities in
\cite{toolkit}-\cite{ciufolani}.

\subsection{The evolution (Raychaudhuri) equations}

We now use the inputs from the previous subsection to rewrite the
evolution equation for $B_{ij}$ as four coupled, nonlinear, first
order equations involving the dependent variables $\theta$,
$\sigma_+$, $\sigma_\times$ and $\omega$. Before that, of course,
we need to write down a general expression for $\ddot \xi^i$. This,
we assume to be (in the linear approximation) of the form
\begin{equation}
\ddot \xi^{i}\, = - K^i_{\,\,j} \, \xi^j -\beta \, \dot{\xi}^i\, , \\
\label{sdform}
\end{equation}
where $K_{ij}$ and $\beta$ represent the stiffness and viscous damping
in the medium, respectively, with $K_{ij}$ of the form 
\begin{equation}
K_{ij}=
\left(
\begin{array}{cc}
k+k_{+}& k_{\times} \\
k_{\times} & k-k_{+} \\
\end{array} \right).
\end{equation}
Using (\ref{sdform}) in (\ref{ddotxi}) leads to
\begin{equation}
\dot{B}^i_{\,\,j}+B^{i}_{\,\,k}B^k_{\,\,j}+K^i_{\,\,j}+\beta B^i_{\,\,j}=0.
\label{dotb}
\end{equation}
Finally, using the definition (\ref{bij}) in (\ref{dotb}), one obtains the
differential evolution (in time) of the ESR as given by the equations:
\begin{equation}
\dot \theta + \frac{1}{2}{\theta^2}+\beta \theta  + 2 (\sigma_{+}^{2} +\sigma_{\times}^{2} -\omega^2) + 2k =0,
\label{theta}
\end{equation}
\begin{equation}
\dot \sigma_{+}+ (\beta +\theta) \, \sigma_{+} + k_{+}=0,
\label{sigma+}
\end{equation}
\begin{equation}
\dot \sigma_{\times}+ (\beta +\theta) \, \sigma_{\times} + k_{\times}=0,
\label{sigmacross}
\end{equation}
\begin{equation}
\dot \omega+ (\beta + \theta)\, \omega=0
.\label{omega}
\end{equation}
The equation for $\theta$ is known as a Ricatti differential equation in 
the mathematics literature. It is nonlinear, first order and its solutions
are known.  
Note that the equations for $\sigma_+$, $\sigma_\times$ and $\omega$ are 
structurally similar. We now look for  analytical solutions of these equations 
and try to identify generic features which appear in the solutions.

\subsection{Analytical solutions}

To find solutions, it is useful to look at specific cases before
we attempt the most general situation. To this end, we consider the following 
cases : 
\begin{enumerate}
\item{ $K_{ij} =0$ and $\beta = 0$,}
\item{$K_{ij}= k\delta_{ij}$ and $\beta =0$,}
\item{$K_{ij}= k\delta_{ij}$ and $\beta \neq 0$.}
\end{enumerate}
The equations (\ref{theta})-(\ref{omega}) for each of the
above cases can be obtained easily, and analytical solutions can
then be derived. We now discuss below each case in detail. \\

\noindent{\bf Case 1:} $K_{ij}=0$ and $\beta=0$ \\

We first consider the following pair
of equations,
\begin{equation}
\dot \theta  +\frac {1}{2} \theta^2 +2I =0
,\label{thetai}
\end{equation}
\begin{equation}
\dot I  + 2\theta I =0,
\label{I}
\end{equation}
where $I =\sigma_{+}^{2} +\sigma_{\times}^{2} -\omega^2$.  It is clear from
the equation for $I$ that its general solution would be of the form
$I=I_0 e^{-2\int \theta (t) dt}$. This means that if $I=0$ at $t=0$, it
would remain zero forever. Further, $I$ can not change sign during
its evolution. 
The scheme for solving equations (\ref{thetai}) and \, (\ref{I})  would be
to replace $\theta$ in (\ref{thetai})  by $-\frac{1}{2}\frac{\dot I}{I}$,
then solve it for $I$ and use it to find $\theta$ from (\ref{I}).
Once $\theta$ is known, the equations for $\sigma_+$, $\sigma_\times$ and
$\omega$ can be solved easily. It is therefore obvious that the solutions to 
the equations (\ref{thetai}) and (\ref{I}) depend on the conditions imposed on $I$.
We use this fact to obtain the different types of solutions.  The
exact solutions for $\theta\, \,, \sigma_{+}, \, \,
\sigma_{\times}$ and $\omega$ with different conditions on $I$ are  given in Table \ref{t1}.
\begin{table}[H]
\small
\begin{center}
\begin{tabular}{|c|c|c|}
\hline\, \, \,  $I$\, \, \, & \, \, \, $\theta$ &\, \, $ \{\, \, \sigma_{+},\,\, \, \sigma_{\times},\, \, \,\omega\, \,  \} \, \, $\\
 \hline &&\\
\, \, \,  $I>0$ \, \, \, & \, \, \,  $ \frac{C \left(D +\frac{Ct}{4}\right)}{2
\left[(D+ \frac{Ct}{4})^2 -16\right]}$\, \, \, \, \,  &  $\frac{\{\, \, E,\, \, \, F,\, \, \, G\, \}}{ \left[(D+ \frac{Ct}{4})^2 -16 \right]}$ \\  
 \hline &&\\\, \, \, $I=0$\, \, \, &  \, \, \,
 $ \frac{\theta_0}{ \left (1+\frac{\theta_0 t}{2} \right)}
$ &$\frac{\{\, \, \sigma_{+ 0}, \, \, \, \sigma_{\times 0},\,\, \,\omega_{ 0}\, \, \}} { \left (1+\frac{\theta_0 t}{2} \right )^2}$  \\ \hline&&\\
\, \, \,  $I<0$ \, \, \, & $ \frac{C \left(D+ \frac{Ct}{4}\right)}{2 \left[(D +\frac{Ct}{4})^2 +16\right]}$& $\frac{\{\, \, E,\, \,F,\,\, G \, \, \}}{ \left[(D + \frac{Ct}{4})^2
 +16\right]}$ \\
\hline
\end{tabular}
\end{center}
\caption {The analytical solutions for Case 1 in two dimensions.} 
\label{t1}
\end{table}
In Table~\ref{t1}, $C$, $D$, $E$, $F$ and $G$ are the integration constants, andwill  be used throughout hereafter in the text to represent the integration constants for  the analytical solutions of other cases as well. Further, we have used $\{ \, \}$ brackets in Table \ref{t1} to express the solutions in a compact way for certain ESR variables which have a common factor. This notation has been
used throughout in the text. For the present  case, the integration constants can be defined in terms of the initial conditions on the ESR variables at $t=0$ using  $D= \frac{2 \theta_{0}}{ \sqrt{ I_0 }}$, where $I_0 =\sigma_{+0}^{2} +\sigma_{\times 0}^{2} -\omega_{0}^2$ (the extra subscript $0$ indicates the initial value of the corresponding ESR variable). The other integration constants can then
be easily written as follows : 
%(i) \noindent $\underline{ when \, I>0} :$
\begin{enumerate}
\item{For $I>0$: 
\begin{eqnarray}
C &=& \, 2 \theta_{0} \left (\frac{D^2- 16}{D}\right ),  \\
\{\, E, \, F, \, G\, \} & =& \{\, \sigma_{+0},\, \sigma_{\times 0},\,  \omega_{0}\, \} \, (D^2 -16).
\end{eqnarray}} 
% \noindent $\underline{ when \, I<0} : $
\item{For $I<0$: 
\begin{eqnarray}
C& =& \, 2 \theta_{0} \left(\frac{D^2+ 16}{D} \right), \\
\{\, E,\, F,\,  G \,\} & =&\{\, \sigma_{+0}, \, \sigma_{\times 0}, \, \omega_{0}\} \,  (D^2 +16),
\end{eqnarray} }
\end{enumerate}
 where $D$ is redefined as $\frac{2 \theta_{0}}{ \sqrt{- I_0 }}$. For $I=0$, the solutions in Table~\ref{t1} are given in terms of the initial conditions on the kinematical 
quantities themselves. 

\begin{figure}
\centerline{\includegraphics[scale=0.5]{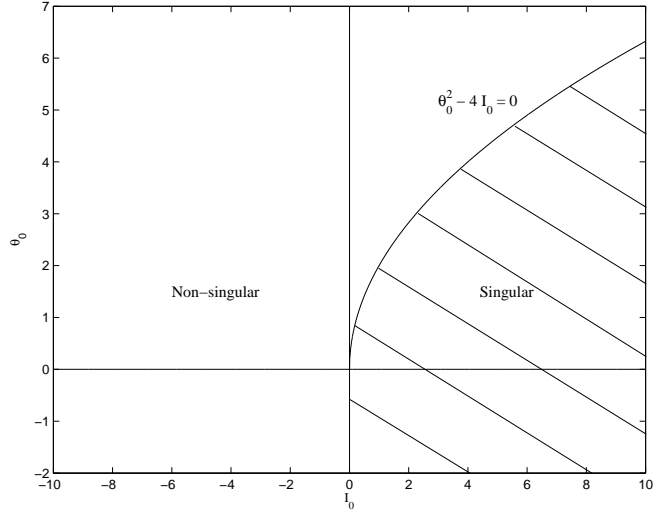}} \caption{
Regions of initial conditions for singular and non-singular solutions in two dimensions.} 
\label{fig3}
\end{figure}
\begin{figure}
\centerline{\includegraphics[scale=0.68]{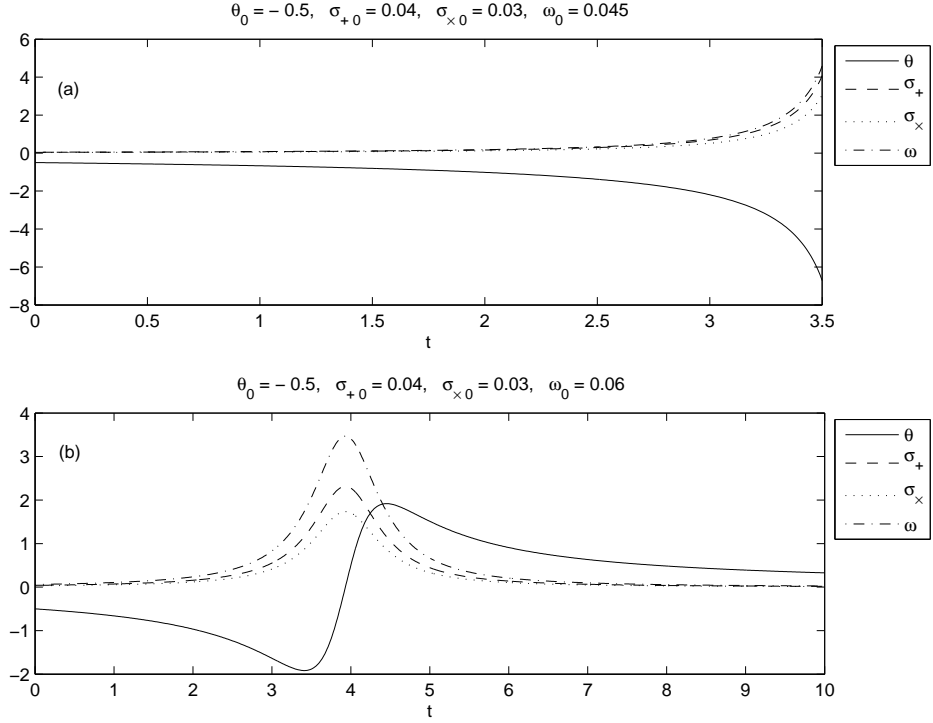}}
\caption{Case 1 in two dimensions: (a) for $I_0>0$, and (b) for $I_0<0$. }
\label{fig1}
\end{figure}
\begin{figure}
\centerline{\includegraphics[scale=0.68]{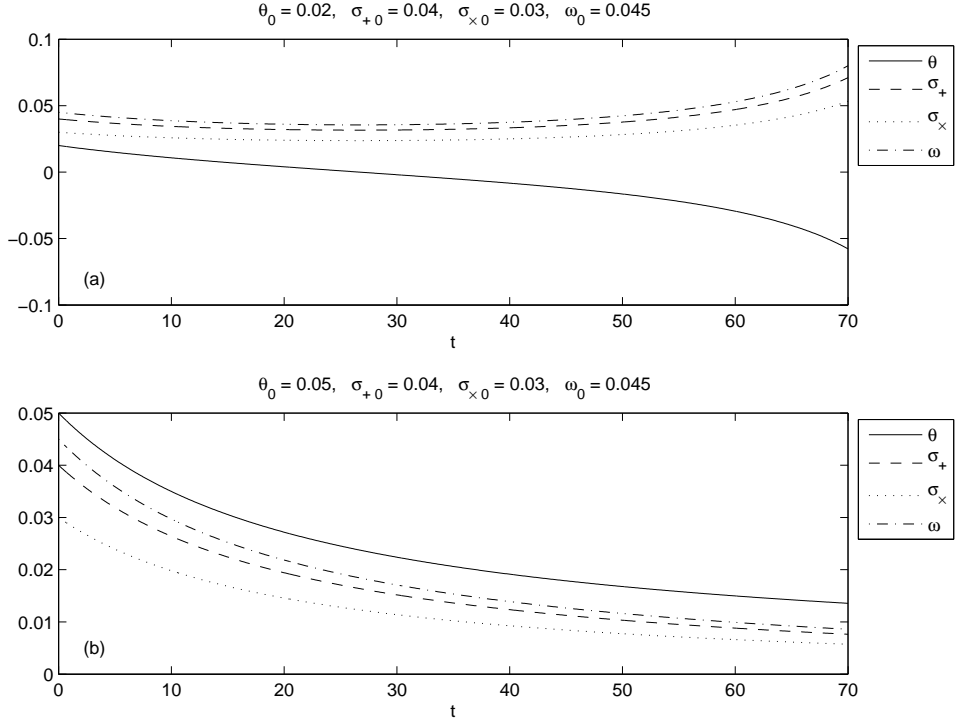}} 
\caption{Case 1 in two dimensions for two different initial values of expansion scalar $\theta_0$. }
\label{fig2}
\end{figure}

We note from Table~\ref{t1} that for $I>0$, the solutions will tend to a singularity in finite time
whenever $D<4$, i.e., $\theta_0^2-4I_0<0$, while for $I=0$, we have a singularity for $\theta_0<0$.
On the other hand, there is no singularity for $I_0>0$. This dependence of the occurrence of a singularity on the initial conditions is summarized on the $I_0$-$\theta_0$ plane in 
Fig.~\ref{fig3}. In Figs.~\ref{fig1} and \ref{fig2}, the evolution of the ESR variables is  shown for different
initial conditions. The initial conditions given in Fig.~\ref{fig1}(a) satisfy $I_0>0$, 
while for the initial condition of Fig.~\ref{fig1}(b), $I_0<0$.
It may be noted that the two initial conditions differ only in $\omega_0$. In the second case 
(i.e., $I_0<0$), the singularity condition $\theta_0-4I_0<0$ can never be satisfied. 
The negative sign of $I_0$ can only occur if $\omega_0$ is large enough 
(i.e., $\omega_0^{2}>\sigma_{+0}^2+\sigma_{\times 0}^2$). Thus,
it may be inferred that there is a critical value of $\omega_0$ that makes the solution non-singular. 
In other words, one can say that presence of a minimum rotation can prevent the appearance of a singularity. 
This is explained further in Section~\ref{visu}. 
In case the amount of rotation present is not large enough (i.e. $I_0>0$), the solution can still 
avoid a singularity if the expansion is large enough to prevent a collapse. This minimum value of
expansion variable is obtained again from the singularity condition. Figure~\ref{fig2} shows the evolution of
the ESR  variables for two values of $\theta_0$ (when $I_0>0$). \\
\noindent{\bf Case 2:} $K_{ij}=k\delta_{ij}$ and $\beta=0$ \\
In this case the equation (\ref{I}) remains same, while (\ref{thetai}) has an additional term $2k$ on the left hand side. Therefore, the form of solution for $I$ will be same as in the Case $1$.  The exact solutions for ESR  variables with different possible conditions are then given in the Table \ref{t3}.
\begin{table}[H]
\small
\begin{center}
\begin{tabular}{|c|c|c|}
\hline $I$& \, \, \, $\theta$ &\, \, $ \{\, \, \sigma_{+},\,\, \, \sigma_{\times},\, \, \,\omega\, \,  \} \, \, $
\\
\hline  $I>0$ &&\\
$ a^2 = 16 +b^2, \, p^2 = 16k$ &$\frac{ p a\, \cos (D + \frac{p t}{2})}{2[a \,
\sin (D + \frac{p t}{2}) + b]}$&$\frac{\{\, \, E,\, \, \, F,\, \, \, G\, \}}{[a \, \sin (D + \frac{p
t}{2}) + b]} $ \\ \hline
&&\\  $I=0$& $2 \sqrt{k} \tan [\sqrt{k} (C-t)]$&$ \{\, \, D,\, \, \, E,\, \, \, F\, \} \sec^2 [\sqrt{k} (C-t)]\, 
$\\ \hline
$I<0$ &&\\ $ a^2 = 16 +b^2, \, p^2 = 16k$ & $\theta =  \frac{ p a\, \sin (D + \frac{p t}{2})}{2[a \, \cos (D +
\frac{p t}{2}) + b]} $&$\frac{\{\, \, E,\, \, \, F,\, \, \, G\, \}}{[a \, \cos (D + \frac{p t}{2}) + b]}$ \\
\hline
\end{tabular}
\end{center}
\caption {The analytical solutions for Case 2 in two dimensions.} \label{t3}
\end{table}
The constants of integration in terms of initial conditions are given as follows
\begin{enumerate}
\item{For $I>0$:
\begin{eqnarray}
b&=& \sqrt{\frac{I_0}{p^2}}
\left( -8 + \frac{2\theta_0^{2}}{I_0} +\frac{p^2}{2I_0}\right ), \label{blesscaseiii} \\
D& =& \tan^{-1} \left(\frac{p-b\sqrt I_0}{2\theta_0}\right),  \\
\{\,E,\, F,\, G\,\}& =& \{\,\sigma_{+0},\,\sigma_{\times 0},\, \omega_{0}\, \}\, (a \, \sin D + b).
\end{eqnarray}}
\item{For $I=0$:
\begin{eqnarray}
C& =& \frac{1}{\sqrt k} \, \tan^{-1} \left (\frac{\theta_0}{2\sqrt {k}} \right ),  \\
\{\,D,\, E,\, F\,\}& =& \frac{\{\,\sigma_{+0},\,\sigma_{\times 0},\,  \omega_{0}\,\}}{\sec^2 (\sqrt{k}\, C)} .
\end{eqnarray}}
\item{For $I<0$:
\begin{eqnarray}
D& =& \tan^{-1} \left(\frac{2\theta_0}{p-b\sqrt{ - I_0}}\right),  \\
\{\, E,\,  F,\, G \,\}& =& \{\, \sigma_{+0},\, \sigma_{\times 0}, \, \omega_{0}\, \} \, (a \, \cos D + b),
\end{eqnarray}}
\end{enumerate}
where $ b$ is same as given in equation (\ref{blesscaseiii}) for $I>0$. In this case, it can be easily noticed that 
all solutions become singular in finite time.\\

\noindent{\bf Case 3:} $K_{ij}=k\delta_{ij}$ and $\beta\ne 0$ \\
For this general case, the equations (\ref{thetai}) and (\ref{I}) are modified as follows
\begin{equation}
\dot \theta  +\frac {1}{2} \theta^2 + \beta \theta  +  2I  + 2k = 0
,\label{btheta}
\end{equation}
\begin{equation}
\dot I  + 2 (\beta +\theta) I =0.
\label{betai}
\end{equation}
The equation (\ref{betai}) has its general solution of the form  $ I=I_0 e^{-2\int [\beta +\theta (t)] dt}$. 
Further, we have various special cases resulting due to different possible conditions on the combination of stiffness and damping. 
The exact solutions for the ESR  variables  are 
calculated and given in the Table~\ref{t4}.
\begin{table}
\small
\begin{center}
\begin{tabular}{|l|c|c|}
\hline \hspace{1.1in} $I$ & $\theta$ & 
$ \{\, \, \sigma_{+},\,\, \, \sigma_{\times},\, \, \,\omega\, \,  \} \, \, $
\\
 \hline  $I>0$&&\\
\hspace{0.1in} $(i)\,\, \, \beta^2>4k$ && \\
\hspace{0.3in}$\bullet \,\, \,b^2 < 16$ 

&$ \frac{p a \, \cosh(D +
\frac{pt}{2})}{2\, [a \sinh(D + \frac{pt}{2}) - b]} \,
  - \beta$&$ \frac{\{\, \, E,\, \, \, F,\, \, \, G \,\}} {[a \sinh(D + \frac{pt}{2}) - b]}$ \\ \, \, \,  $a^2=16- b^2, \, \, \, 
p^2 =4(\beta^2 -4k)\, \, \, $&&\\
\hspace{0.3in}$\bullet \,\, \, b^2=16$& $\frac{p \exp(D + \frac{p t}{2})}{2\,[\exp(D
+ \frac{p t}{2}) -b]} \, -\beta$&$\frac{\{\, E,\, \, F, \,  \, G \, \}}{[\exp (D + \frac{pt}{2}) -b]}$\\&&\\
\hspace{0.3in}$\bullet \,\, b^2>16, \, \, a^2=b^2-16$&$\frac{p a \, \sinh (D + \frac{pt}{2})}{2\,[a
\, \cosh(D + \frac{pt}{2}) - b]} \, - \beta$
&$\frac{\{\, \, E,\, \, \, F,\, \, \, G\}}
{[a \, \cosh(D + \frac{pt}{2})- b]}$\\&&\\
\hspace{0.1in}$(ii)\,\beta^2=4k$& \, \, \,  $ \frac{C (\frac{D}{2} +
\frac{Ct}{4})}{2\, [(\frac{D}{2} + \frac{Ct}{4})^2 -16]} \,
-\,\beta$&$\frac{\{\, \, E,\, \, \, F,\, \, \, G\}}{[(\frac{D}{2} + \frac{Ct}{4})^2
-16]}$\\ &&\\
\hspace{0.1in}$(iii)\, \beta^2<4k$ &$ \frac{ p a\, \cos (D + \frac{p t}{2})}{2 \, [a
\, \sin (D + \frac{p t}{2}) + b]}  \,-\, \beta$&$\frac{\{\, \, E,\, \, \, F,\, \, \, G\}}{[a
\, \sin (D + \frac{p t}{2}) + b]}$ \\\, \, \,  $a^2=16+b^2,\,\, \,  p^2 = 4(4k-\beta^2)\, \, \, $&&\\
 \hline
$I=0$&&\\
\hspace{0.1in} $(i)\,\,  \beta^2>4k, \, \, a^2 = \beta^2-4k$& $ \,\frac{a\,[\exp(at) +\exp{2C}\,]}{\exp (at) -\exp(2C)}
 \, -\beta$&$\frac{\{\, \, D,\, \, \, E,\, \, \, F\}\exp(at)}{[\exp (at) -\exp(2C)]^2}$ \\ &&\\ \hspace{0.1in}$(ii)\,\beta^2=4k$& $\frac{2}{(t-C)}-\,
\beta$&$\frac{\{\, \, D,\, \, \, E,\, \, \, F\}}{(t-C)^2}$ \\
&&\\
\hspace{0.1in}$(iii) \beta^2<4k, \, \, a^2 = 4k-\beta^2 $& $a \tan \frac{a}{2}(C-t)$&$\{\, D,\, E,\,  F \, \} \sec^2 \frac
{a}{2}(C-t)$ \\  \hline
$I<0$&&\\
\hspace{0.1in}$(i) \beta^2>4k$ &$ \frac{p a \sinh (D + \frac{pt}{2})}{2 \,[a \cosh
(D + \frac{pt}{2}) + b]} \, - \beta $&$\frac{\{\, \, E,\, \, \, F,\, \, \, G\}} { [a \cosh (D + \frac{pt}{2}) + b]}$ \\ $a^2=16 + b^2, \, p^2 = 4(\beta^2 -4k)$&&\\
\hspace{0.1in}$(ii)\,\beta^2=4k$& \, \, \,  $ \frac{C (\frac{D}{2} +
\frac{Ct}{4})}{2 \, [(\frac{D}{2} +\frac{Ct}{4})^2 +16]} \,
-\,\beta$&$\frac{\{\, \, E,\, \, \, F,\, \, \, G\}}
{[(\frac{D}{2} + \frac{Ct}{4})^2 +16]}$\\ &&\\
\hspace{0.1in}$(iii)\,\,  \beta^2<4k$& $  \frac{ p a\, \cos (D + \frac{p
t}{2})}{2 \, [a \, \sin (D + \frac{p t}{2}) + b]}
  \,-\, \beta$&$
\frac{\{\, \, E,\, \, \, F,\, \, \, G\}}{[a \, \sin (D + \frac{p t}{2}) + b]}$ 
\\ \, \, 
$a^2 = b^2 -16 \,, \, \, p^2 = 4(4k-\beta^2)\, \, $&&\\ \hline
\end{tabular}
\end{center}
\caption {The analytical solutions for Case 3 in two dimensions.} \label{t4}
\end{table}
The integration constants used in Table \ref{t4} may then  be defined in the following form
\begin{enumerate}
\item{For $I>0$:
\begin{equation}
b= \sqrt{\frac{I_{0}}{p^2}} \left( -8 +\frac{2(\beta
+\theta_0)^{2}}{I_0} -\frac{p^2}{2I_0}\right ),\label{b2less16}
\end{equation}
\begin{enumerate}
\item{$\beta^2 > 4k$ \\
\noindent{$\bullet$ $b^2<16$,} 
\begin{eqnarray}
D& =& \tanh^{-1} \left[\frac{p+b\sqrt I_0}{2(\beta
+\theta_0)}\right], \\
 \{\,E,\, F,\, G\,\}& =& \{\, \sigma_{+0},\, \sigma_{\times 0},\, \omega_{0}\,\}\, (a \, \sinh D - b ).
\end{eqnarray}
\noindent{$\bullet$ $b^2=16$, } 
\begin{eqnarray}
D& =& \ln \left [\frac{2(\beta +
\theta_0)}{\sqrt I_0} \right ],  \\
\{\,F,\, F,\, G\,\}& =& \{\,\sigma_{+0},\,\sigma_{\times 0},\, \omega_{0}\,\}\, [1-\exp(C)]^2 .
\end{eqnarray}
{$\bullet$ $b^2>16$}  
\begin{eqnarray}
D &=& \tanh^{-1} \left[\frac{2(\beta
+\theta_0)}{p+b\sqrt I_0}\right],   \\
\{\,E,\, F,\, G \,\} &=& \{\,\sigma_{+0},\, \sigma_{\times 0},\, \omega_{0} \,\}\, \, (a \, \cosh D - b ).
\end{eqnarray}}
\item{$ \beta^2 =4k$
\begin{eqnarray}
D&=& \frac {4 (\beta +\theta_{0})}{\sqrt {I_0}},   \\
C& =&  (\beta +\theta_{0})\,  \frac{(D^2- 64)}{D}, \\
\{\,E,\, F,\, G\,\}& =& \{\, \sigma_{+0},\,\sigma_{\times 0},\, \omega_{0}\,\}\, \frac{(D^2-64)}{4}.
\end{eqnarray}
}
\item{$ \beta^2 < 4k$ 
\begin{eqnarray}
b&=& \sqrt{\frac{I_{0}}{p^2}} \left( -8 + \frac{2(\beta
+\theta_0)^{2}}{I_0} +\frac{p^2}{2I_0}\right ),
 \\
D &= &
\tan^{-1} \left[\frac{p-b\sqrt I_0}{2(\beta +\theta_0)}\right], \\
\{\,E, \, F,\,  G\, \}& =& \{\, \sigma_{+0},\, \sigma_{\times 0},\,  \omega_{0}\, \}\, (a \, \sin D + b).
\end{eqnarray}}
\end{enumerate}
}
\item{For $I=0$: 
\begin{enumerate}
\item{$ \beta^2 > 4k$ 
\begin{eqnarray}
\exp(2C) &=& \frac{\beta +
\theta_0 -a}{\beta + \theta_0 +a},  \\
 \{\, D,\, E,\, F \,\} &=& \{\, \sigma_{+0},\,\sigma_{\times 0},\, \omega_{0} \,\}\, [\, 1-\exp(2C)\, ]^2 .
\end{eqnarray}}
\item{ $ \beta^2 =4k$
\begin{eqnarray}
C&=& \,\frac{-2}{\beta+\theta_0}, \\
\{\, D,\,  E,\,  F\, \}& =& \{\, \sigma_{+0},\, \sigma_{\times 0},\, \omega_{0}\, \}\,C^2.
\end{eqnarray}}
\item{$ \beta^2 < 4k$ 
\begin{eqnarray}
C &=& \frac{2}{a} \, \tan^{-1} \left (\frac{\theta_0}{a} \right ),  \\
\{\, D,\,  E,\,  F \, \}& =& \frac{\{\,\sigma_{+0},\,\sigma_{\times 0},\,\omega_{0} \, \}}{\sec^2 \frac{aC}{2}} .
\end{eqnarray}}
\end{enumerate}
}
\item{For $I<0$: 
\begin{enumerate}
\item{$\beta^2 > 4k$
\begin{eqnarray}
b&=& \sqrt{\frac{- I_{0}}{p^2}} \left( -8 +  \frac{2
(\beta+\theta_0)^{2}}{I_0} - \frac{p^2}{2I_0}\right ),  \\
 D& =&
\tanh^{-1} \left[\frac{2(\beta +\theta_0)}{p-b\sqrt {- I_0} }\right], \\
\{\, E,\,  F,\,  G\, \} &=& \{\, \sigma_{+0},\, \sigma_{\times 0},\, \omega_{0}\, \}\, (a \, \cosh D +b).
\end{eqnarray}}
\item{$ \beta^2 =4k$, 
\begin{eqnarray}
D&=& \frac{4(\beta +\theta_{0})}{\sqrt{- I_0}},  \\
C &=&  (\beta +\theta_{0})\,  \frac{(D^2+ 64)}{D}, \\
\{\, E,\,  F,\,  G\, \}& =& \{\, \sigma_{+0}, \, \sigma_{\times 0},\,  \omega_{0}\, \} \,\frac {(D^2+64)}{4}.
\end{eqnarray}}
\item{$ \beta^2 < 4k$
\begin{eqnarray}
b&=& \sqrt{\frac{-I_{0}}{p^2}} \left( 8 -\frac{2(\beta
+\theta_0)^{2}}{I_0} -\frac{p^2}{2I_0}\right)  \\
 D& =& \tan^{-1}
\left[\frac{p-b\sqrt{- I_0}}{2(\beta +\theta_0)}\right], \\
\{\, E,\, F,\, G\, \}& =& \{\, \sigma_{+0},\, \sigma_{\times 0}, \, \omega_{0} \, \}\, (a \, \sin D + b).
\end{eqnarray} }
\end{enumerate}}
\end{enumerate}
In this case, the regions of the ESR  variable space of initial conditions that lead to solution
singularity in finite time is more complex. We identify first some classifying features of the
solutions presented in Table~\ref{t3}. For $I_0>0$ and $\beta^2<4k$, the solution is always singular,
while for $\beta^2\ge 4k$ the solution is non-oscillatory. These non-oscillatory solutions may
or may not be singular depending on further conditions. For example, for $\beta^2>4k$ and $b^2<16$,
we have singularity when $\sinh D<b/a$. For $I_0=0$ and $\beta^2>4k$, the solution has finite
time singularity when $\beta+\theta_0+a<0$, as can be easily checked. The remaining sub-cases in 
Table~\ref{t4} can be analysed similarly. It may be noted that, for $I_0<0$, allsolutions are non-singular, and we have an oscillatory solution only when $\beta^2<4k$.
\begin{figure}
\centerline{\includegraphics[scale=0.7]{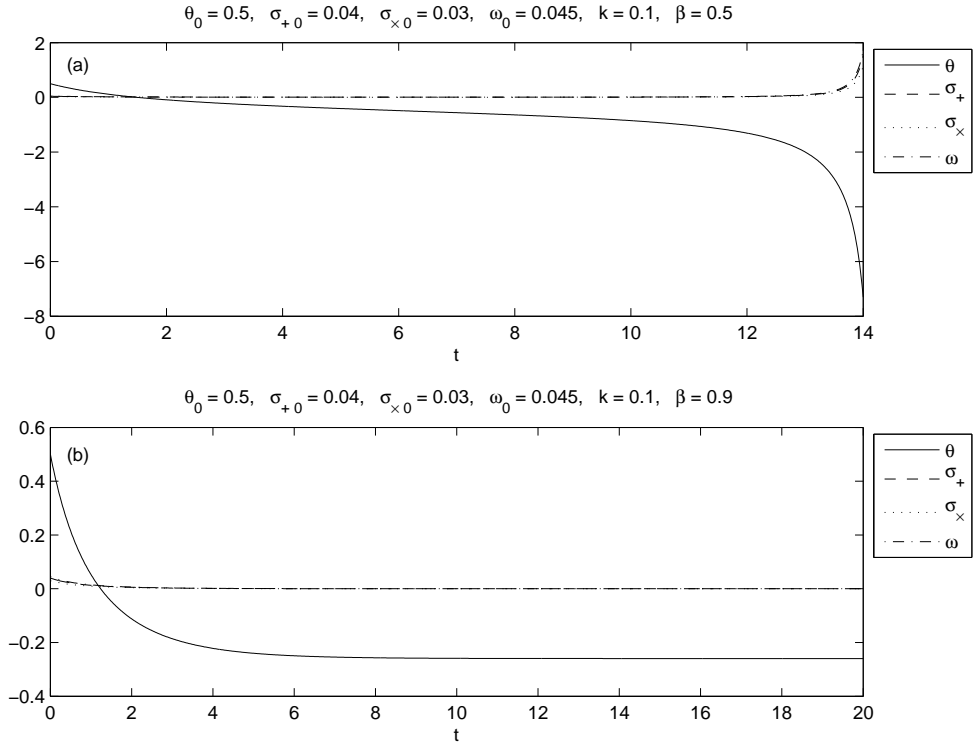}}
\caption{Case 3 in two dimensions: role of damping in suppression of singularity.} 
\label{fig8}
\end{figure}
\begin{figure}
\centerline{\includegraphics[scale=0.7]{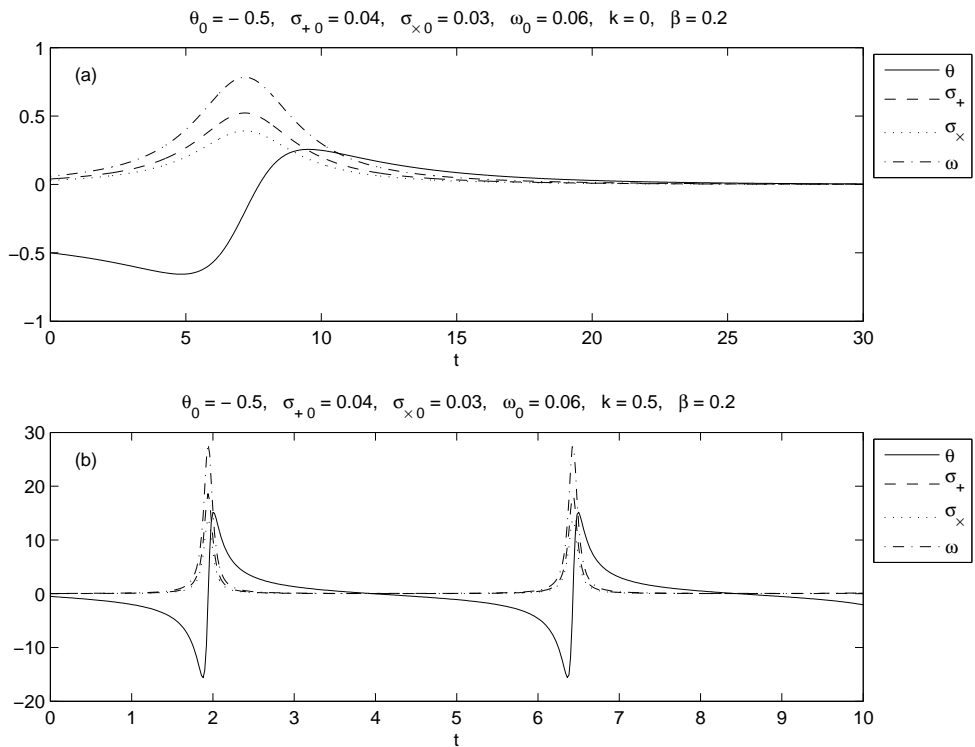}} 
\caption{Case 3 in two dimensions: effect of stiffness in generation of oscillations.} 
\label{fig5}
\end{figure}
The evolution of the ESR variables have been computed from the analytical solutions and 
presented in Figs.~\ref{fig8} and \ref{fig5}.
For $I_0>0$, the effect of damping in avoiding solution singularity is observed in Fig.~\ref{fig8}. The existence of
singularity is evident from Fig.~\ref{fig8}(a) for $\beta^2<4k$, while there is no singularity in Fig.~\ref{fig8}(b) for 
the  initial conditions which satisfy $\beta^2>4k$. It is
interesting to note that for $\beta^2>4k$, the expansion variable settles to a negative value implying that
the medium contracts indefinitely, however without the appearance of a 
singularity.  On the other hand, for $I_0<0$, the presence of stiffness produces oscillations as
seen in Fig.~\ref{fig5}. However, there is a minimum value of stiffness which induces oscillatory 
behaviour and is given by $k>\beta^2/4$. 

\subsection{Visualisation of deformations of the medium}\label{visu}

The deformations of a medium in two dimensions may be easily visualised
by considering the deformations of, say, a square element defined by four points of
the medium. For such an element, we compute the
deformation vectors from the central geodesic (taken as the centroid of the square)
to the four corner points of the square at successive time steps. Thus, the four points
define the shape of the element at any time instant. 

\begin{figure}
\centerline{\includegraphics[scale=0.25]{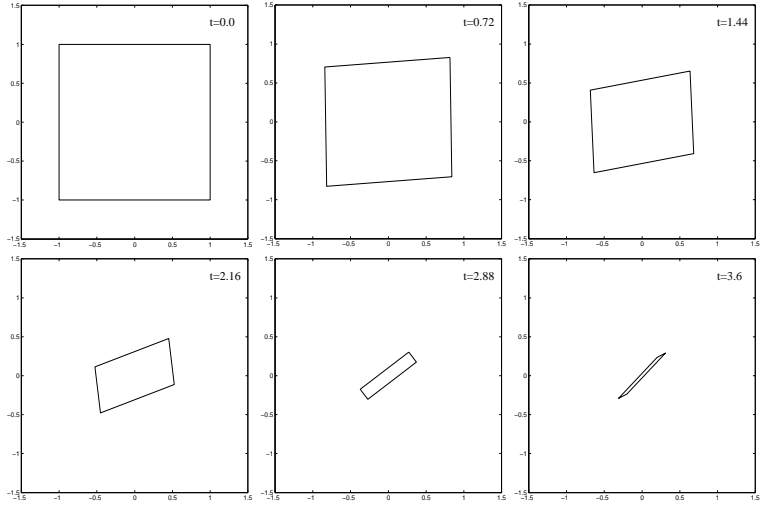}}
\caption{Sequence of deformed configurations of a square element in two dimensions for initial
conditions given in Fig.~\ref{fig1}(a).} \label{seq1}
\end{figure}
Let us consider the first case, with initial conditions as given in Fig.~\ref{fig1}(a). The evolution of 
ESR  variables
in Fig~\ref{fig1}(a) clearly shows a singular behaviour in finite time. In order to understand this behaviour in terms of
the deformations of the medium, we have plotted the successive shapes of an initially square element
in Fig.~\ref{seq1}. It is observed that, as the element shrinks, its area approaches zero. However,  
when $I_0>0$, the rate of change of area $\dot{A}\ne 0$ as $A \rightarrow 0$. 
The expansion variable  in two dimension has the interpretation of area strain rate, i.e., 
$\theta=\dot{A}/A$ (see \cite{toolkit}). Hence $\theta=\dot{A}/A\rightarrow -\infty$, 
signifying a singularity. This can be easily seen as follows.  
Using this definition in (\ref{theta})-(\ref{omega}) (with
$K_{ij}=0$ and $\beta=0$), one can express 
$\dot{A}=\sqrt{4 I_0/A_0^2+ P  A}$, where $A_0$ is
the initial area of the element, and $P$ is a constant. In the limit $A\rightarrow 0$, $\dot{A}\rightarrow (2/A_0)\sqrt{I_0}$. 
\begin{figure}
\centerline{\includegraphics[scale=0.25]{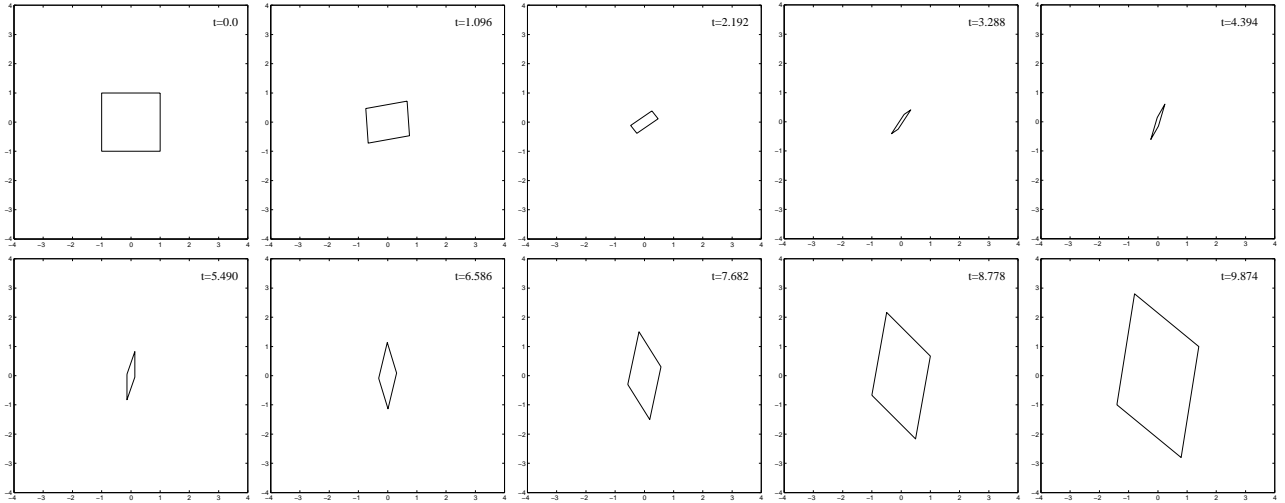}}
\caption{Sequence of deformed configurations of a square element in two dimensions for initial
conditions given in Fig.~\ref{fig1}(b).} \label{seq2}
\end{figure}
\begin{figure}
\centerline{\includegraphics[scale=1.2]{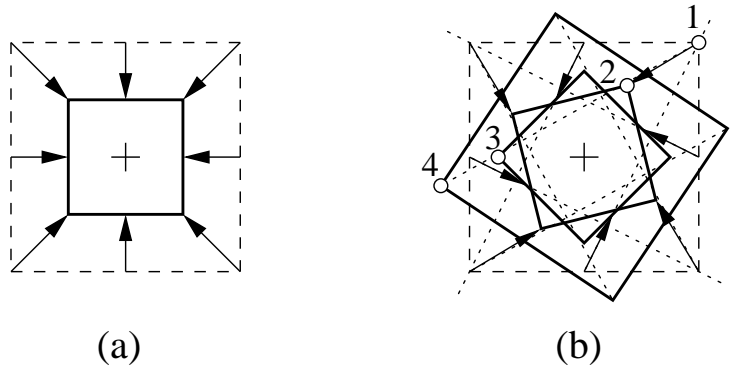}}
\caption{Schematic representation of deformation of a square element in two dimensions for
negative expansion (a) without rotation, and (b) with rotation.} \label{contsh}
\end{figure}
For the initial conditions of Fig.~\ref{fig1}(b),
the successive shapes of the square element is shown in Fig.~\ref{seq2}. It may be noted in this figure
that the square initially shrinks when $\theta<0$, but later expands when $\theta>0$. The presence
of sufficient rotation helps in avoiding the singularity situation, as 
explained through Fig.~\ref{contsh} (for no shear). In the case when the expansion is negative, and shear and
expansion  variables are zero, the element collapses isotropically, as shown in Fig.~\ref{contsh}(a).
However, when a slight rotation is added to the collapsing element, singularity is avoided,
as observed in Fig.~\ref{contsh}(b). It may be noted in Fig~\ref{contsh}(b) that
when a corner point of the square (marked by a circle) is closest to the central geodesic, rapid
rotation of the element is observed. In case shear is also present, there is a minimum initial
rotation required to avoid a singularity. This minimum value of initial rotation is that which makes
$I_0<0$. Finally, one may observe that, due to the 
positive {\em tail} of the expansion variable $\theta$ in Fig.~\ref{fig1}(b), the element in 
Fig.~\ref{seq2} expands even as shear and rotation of the element cease.
\begin{figure}
\centerline{\includegraphics[scale=0.25]{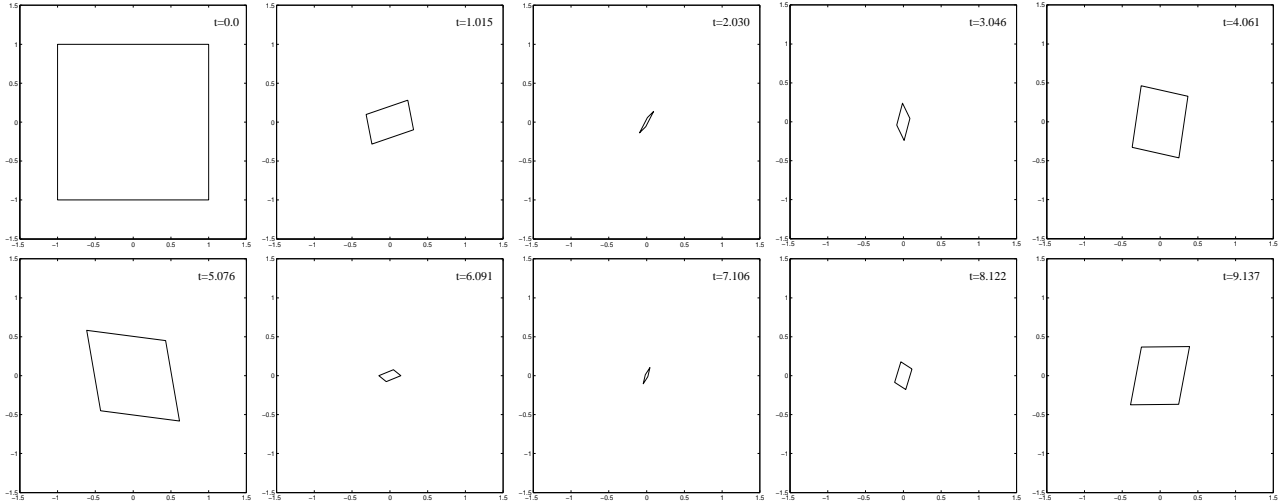}}
\caption{Sequence of deformed configurations of a square element in two dimensions for initial
conditions given in Fig.~\ref{fig5}(b).} \label{seq3}
\end{figure}
For the case of initial conditions shown in Fig.~\ref{fig5}(b), where
the response of all the variables is oscillatory, the deformations of the medium is visualised in Fig.~\ref{seq3}.
It is interesting to note that, due to the presence of damping in this case, the element
progressively shrinks in size. Further, periodic occurrence of rapid shear and rotation
about the points, where the expansion variable changes sign is observed. In case the damping
is absent, no shrinking of the area will take place.

%\subsection{Remarks on generic features}

\section{Deformations in three dimensional media}

\subsection{Expansion, rotation and shear}

The basic scheme of analysing the deformations in two dimensions extend to
three dimensions easily. However, we now have nine independent variables here
(one for expansion, five for shear and three for rotation). So, the
level of complicacy is certainly higher as compared to the
two dimensional analysis in the previous section.

As before, we begin by writing down the general expression for $B_{ij}$:
\begin{equation}
B_{ij} = \frac{1}{3}{\theta} \, \delta_{ij} +\sigma_{ij} + \omega_{ij}, \label{bij3d}
\end{equation}
where the expansion $(\theta)$, shear $(\sigma_{ij})$  and rotation $(\omega_{ij})$ are given,
\begin{equation}
\frac{1}{3}\theta\delta_{ij}=
\left(
\begin{array}{ccc}
\frac{1}{3}\theta & 0&0\\
0&  \frac{1}{3}\theta&0\\
0&0&\frac{1}{3}\theta\\
\end{array} \right),~~~~
\sigma_{ij}=
\left(
\begin{array}{ccc}
\sigma_{11} & \sigma_{12}&  \sigma_{13} \\
\sigma_{12} & \sigma_{22}&  \sigma_{23}\\
 \sigma_{13} & \sigma_{23}&-(\sigma_{11} + \sigma_{22})\\
\end{array} \right), ~~~~
\omega_{ij}=
\left(
\begin{array}{ccc}
0 & -\omega_3 & \omega_2 \\
\omega_3& 0 & -\omega_1\\
-\omega_2&\omega_1&0\\
\end{array} \right).
 \label{ctrans3d}
\end{equation}
\subsection{The evolution(Raychaudhuri) equations}
\noindent The evolution equations (i.e. Raychaudhuri equations) for the
expansion, shear and rotation can be then obtained  in three dimensions in
flat space. The nine coupled nonlinear equations for $\theta$ (expansion), $\sigma_{11}, \, \sigma_{12},\,\sigma_{22}\,,\sigma_{13}$,\,$\sigma_{23}$ (shear) and $\omega_{12},\,\omega_{13},\,\omega_{23}$ (rotation) are derived (for the
case when  $K_{ij}= k\delta_{ij}$ and $\beta \neq 0$) in the following form,
\begin{equation}
\dot \theta + \frac{1}{3}{\theta^2}+\beta \theta  + 2 (\sigma_{11}^{2} + \sigma_{22}^{2} + \sigma_{12}^{2} +\sigma_{13}^{2} + \sigma_{23}^{2} + \sigma_{11} \sigma_{22} -\omega_1^{2}-\omega_2^{2}-\omega_3^{2}) +3k=0
,\label{theta3}
\end{equation}
\begin{equation}
\dot \sigma_{11}+ (\beta + \frac{2}{3}\theta) \, \sigma_{11} + \frac{1}{3} (\sigma_{11}^{2} + \sigma_{12}^{2} + \sigma_{13}^{2} - \omega_2^{2} - \omega_3^{2})  - \frac{2}{3} ( \sigma_{22}^{2} + \sigma_{23}^{2} + \sigma_{11} \sigma_{22} -\omega_1^2) =0
,\label{sigma11}
\end{equation}
\begin{equation}
\dot \sigma_{22}+ (\beta + \frac{2}{3}\theta) \, \sigma_{22} +  \frac{1}{3} (\sigma_{21}^{2} + \sigma_{22}^{2} + \sigma_{23}^{2}-\omega_1^{2} -\omega_3^{2})  - \frac{2}{3} ( \sigma_{11}^{2} + \sigma_{13}^{2} + \sigma_{11} \sigma_{22}-\omega_2^2 ) =0
,\label{sigma22}
\end{equation}
\begin{equation}
\dot \sigma_{12}+ (\beta + \frac{2}{3}\theta + \sigma_{11} + \sigma_{22}) \, \sigma_{12} +  \sigma_{13} \sigma_{23} +  \omega_1 \omega_2 =0
,\label{sigma12}
\end{equation}\begin{equation}
\dot \sigma_{13}+ (\beta + \frac{2}{3}\theta -\sigma_{22}) \, \sigma_{13} + \sigma_{12} \sigma_{23} + \omega_1\omega_3 =0, \label{sigma13}
\end{equation}
\begin{equation}
\dot \sigma_{23}+ (\beta + \frac{2}{3}\theta -\sigma_{11}) \, \sigma_{23} +  \sigma_{13} \sigma_{12} + \omega_2 \omega_3 =0, \label{sigma23}
\end{equation}
\begin{equation}
\dot \omega_1 + (\beta + \frac{2}{3} \theta - \sigma_{11}) \, \omega_{1} - \omega_2 \sigma_{12} - \omega_3 \sigma_{13} =0
,\label{omega1}
\end{equation}
\begin{equation}
\dot \omega_2 +(\beta + \frac{2}{3}\theta -\sigma_{22}) \, \omega_{2} - \omega_1 \sigma_{12} - \omega_3 \sigma_{23} =0
,\label{omega2}
\end{equation}
\begin{equation}
\dot \omega_3 + (\beta + \frac{2}{3} \theta + \sigma_{11} + \sigma_{22}) \, \omega_{3} -  \omega_1 \sigma_{13} -  \omega_2 \sigma_{23}=0
.\label{omega3}
\end{equation}
It is possible to note some structural similar between subsets of the
above set of equations. The equations for $\sigma_{11}$ and $\sigma_{22}$
are similar and we can obtain one from the other by interchanging and
replacing indices appropriately. In the same way, the three equations
for the off--diagonal components of the shear, i.e. $\sigma_{12}$, $\sigma_{13}$and $\sigma_{23}$ are also similar in structure. A structural similarity
can also be noted for the three equations representing the evolution of
the components of the rotation.  

\subsection{Analytical solutions: some special cases}

In the following, we solve the above equations under the following assumptions: 
$\sigma_{13}=\sigma_{23}=0 $,  $ \omega_1=\omega_2=0$ and  $\omega_3=\omega$. 
We now use the definition $J= \sigma_{12}^{2} -\omega^{2}$.
It may be checked that these assumptions are consistent, i.e., they remain satisfied for all
times if they are satisfied at $t=0$. Now we have only five variables representing the ESR, namely, $\theta$,
$\sigma_{11}$, $\sigma_{12}$, $\sigma_{22}$ and $\omega$. The evolution equations for these along with the procedure followed to solve the equations are given in the Appendix B. 
We have obtained analytical solutions for the following two cases:
\begin{enumerate}
\item{For $K_{ij} =0$ and $\beta = 0$,}
\item{For $K_{ij}= k\delta_{ij}$ and $\beta =0$.}
\end{enumerate}
These are now discussed in detail. \\

\noindent{\bf Case 1}: $K_{ij} =0$ and $\beta = 0$ \\
\begin{enumerate}
\item{$J> 0$: \\
The exact solutions for $\theta$, $\sigma_{11}$, $\sigma_{12}$, $\sigma_{22}$ and $\omega$ are given below,
\begin{eqnarray}
\theta &=& \,\frac{2 E\,(D +  E t)}{[(D +  E t)^2-1]} \, + \, \frac{1}{(t-C)},  \\
\{\, \sigma_{11}, \sigma_{22}\,\} & =& \frac{1}{2}\left[\frac{\theta}{3} -\frac{1}{(t-C)} + \frac{\{\, F, -F\,\}}{[(D+   E t)^2-1]} \right ],  \\
\{\sigma_{12},\,\,\omega\}& =& \frac{\{G,\,\,H\}}{[(D +   E t)^2-1]},   
\end{eqnarray}
where $C$ and $D$ are given in terms of initial conditions as follows,
\begin{eqnarray}
C&=&   \frac{3}{ 3\,[\,  (\sigma_{11})_0 + (\sigma_{22})_0\, ]\,  - \, \theta_0 },  \label{c3digt0} \\
D&=&  \, \frac{{2}\, \theta_0 + 3 (\sigma_{11})_0 +
3 (\sigma_{22})_0}{ 3\, \sqrt{ [\, (\sigma_{11})_0 -(\sigma_{22})_0 \, ]^2 + 4 J_0}}.
\label{d3digt0}
\end{eqnarray}
The integration constants $E$, $F$, $G$ and $H$ can be written in terms of $C$ and $D$ as
\begin{eqnarray}
E&=&   \frac{(1+ \theta_0 \, C ){(D^2-1)}}{2CD},  \\
F&= &  \, [\, (\sigma_{11})_0 - (\sigma_{22})_0 \,\, ](D^2-1),   \\
\{\, G,\, H \, \}&=&   \{ \, (\sigma_{12})_0, \,\omega_0 \, \}\, (D^2-1).  
%H &=& (\omega)_0 \, (D^2-1). 
\end{eqnarray}
These six integration constants (i.e. $C$-$H$) satisfy the condition 
\begin{eqnarray}
2(D^2-1)^2 J_0 + {F^2} -4 E^2 = 0. 
\end{eqnarray}}
\item{$ J= 0$:
\begin{eqnarray}
\theta& =& \frac{1}{(t-C)} + \frac{1}{(t-D)} + \frac{1}{(t-E)},  \\
\sigma_{11}& =& \frac{1}{3}\left[\frac{2}{(t-D)}- \frac{1}{(t-C)}-\frac{1}{(t-E)} \right],  \\
\sigma_{22}& =&  \frac{1}{3}\left[\frac{2}{(t-E)}- \frac{1}{(t-C)}-\frac{1}{(t-D)} \right],  \\
\{\sigma_{12},\,\,\omega\}& =& \frac{\{\, F,\,G\,\}}{(t-D)\,  (t-E)},  
%\omega & =& \frac{G}{(t-D) (t-E)},  
\end{eqnarray}
where $D$, $E$, $F$ and $G$ are given below in terms of initial conditions while $C$ is same as in (\ref{c3digt0}),
\begin{eqnarray}
D&=&  \,- \frac{3}{3(\sigma_{11})_0 + \theta_0}, \\
E&= & - \frac{3}{3(\sigma_{22})_0 + \theta_0 },  \\
\{\, F,\,G\, \}&=&  \{\, (\sigma_{12})_0,\,\, \omega_0 \, \}\,  D \,  E. 
%G&=&  (\omega)_0\,  D\,  E.
\end{eqnarray} 
}
\item{For $J< 0$:
\begin{enumerate}
\item{$[\, (\sigma_{11})_0 -(\sigma_{22})_0 \, ]^2 < 4 \, \vert J_0 \vert $ \\
\begin{eqnarray}
\theta& =& \frac{2 E\,(D + E t)}{[(D +  E t)^2+1]} \, + \, \frac{1}{(t-C)},
 \\
\{\, \sigma_{11}, \sigma_{22}\, \}& =& \frac{1}{2}\left[\frac{\theta}{3} -\frac{1}{(t-C)} + \frac{\{\,F, -F\,\}}{[(D+  E t)^2+1]} \right ],
 \\
\{\, \sigma_{12},\,\omega \, \} &=& \frac{\{G,\,\,H\}}{[(D +  E t)^2+1]}, 
%\, \, \, \omega  = \frac{H}{[(D  + E t)^2+1]}\,,
\end{eqnarray}
where  $C$ is same as in (\ref{c3digt0}) and $D$ are given in terms of initial conditions as follows
\begin{eqnarray}
D&=&  \frac{2 \theta_0 + 3 (\sigma_{11})_0 + 3
(\sigma_{22})_0 }{3\, \sqrt{ [\, (\sigma_{11})_0 -(\sigma_{22})_0 \, ]^2 - 4 J_0}},\label{djless0}
 \\
E&=&   \frac{(1+ \theta_0 \, C ){(D^2+1)}}{2CD}, \\
F&=&  [\, (\sigma_{12})_0 -(\sigma_{22})_0\,]\,  \, (D^2+1),  \\
\{\, G,\, H \, \}&=& \{ \, (\sigma_{12})_0,\,\,\omega_0 \, \} \, (D^2+1). 
%\, \, \,  \,H= (\omega)_0 (D^2+1).
\end{eqnarray}
These integration constants satisfy the following condition, 
\begin{eqnarray}
2(D^2+1)^2\, J_0 + {F^2} - 4 E^2 = 0. 
\end{eqnarray}}

\item{$[\,(\sigma_{11})_0 -(\sigma_{22})_0 \, ]^2 = 4 \vert J_0 \vert $\\
\begin{eqnarray}
\theta &=& \frac{1}{(t-C)} + \frac{2}{(t-D)},
 \\
\{\sigma_{11}, \, \sigma_{22}\} &=& \frac{1}{2}\left[\frac{\theta}{3}- \frac{1}{(t-C)} +\frac{ \{ \,E, \, -E\, \}}{(t-D)^2} \right],\\
\{\, \sigma_{12}, \,\omega \, \}& =& \frac{\{\, F,\,G \, \}}{(t-D)^2 },
\end{eqnarray}
where $C$ is same as in (\ref{c3digt0}) while other integration constants are defined as below,
\begin{eqnarray}
D&=&  - \frac{2C}{(C \theta_0 + 1)}, \\ 
E&=&  \{ \, (\sigma_{11})_0 - (\sigma_{22})_0 \, \}  \, D^2,  \\
\{ \, F,\,G \,\}&=&  \{\, (\sigma_{12})_0,\,\,\omega_0 \, \}\,  D^2. 
\end{eqnarray}}

\item{$[\,(\sigma_{11})_0 -(\sigma_{22})_0\,]^2 > 4 \vert J_0 \vert $\\
The solutions remain same as for the case of $J>0$ with the integration 
constant $D$ given by (\ref{djless0}).
}
\end{enumerate}}
\end{enumerate}

It may be noted that the regions in the space of initial conditions that lead to
finite time singularity of the solutions are more complex in the three dimensional case. For example,
for $J_0>0$, the solution has singularity when either $C>0$ (defined in (\ref{c3digt0})) 
or $D<1$ (defined in (\ref{d3digt0})). On the other hand, for $J_0<0$, we have singularity only
when $C>0$ (defined in (\ref{c3digt0})).
\begin{figure}
\centerline{\includegraphics[scale=0.7]{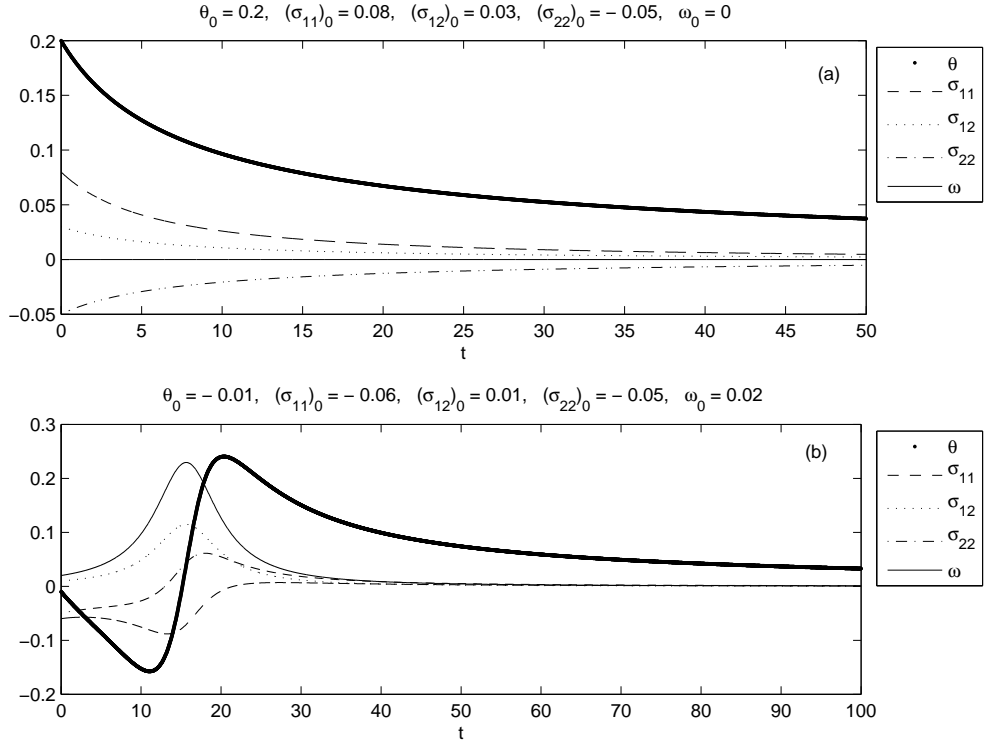}}
\caption{Case 1 in three dimensions: similarity of solutions as also observed in two dimensions. } 
\label{fig11}
\end{figure}
\begin{figure}
\centerline{\includegraphics[scale=0.7]{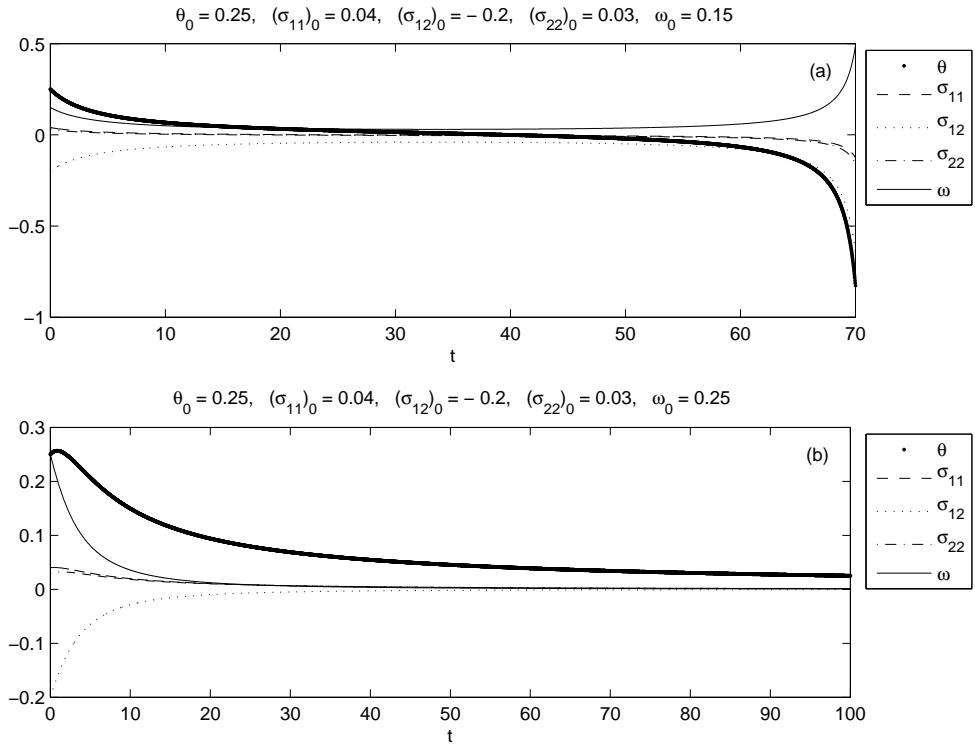}}
\caption{Case 1 in three dimensions: effect of rotation in avoiding singularity.} \label{fig12}
\end{figure}
It may be observed from (\ref{c3digt0}) that $C=-1/B_{33}$. One can interpret $B_{33}$ as the rate of 
expansion/collapse (depending on whether $B_{33}>0$ or $B_{33}<0$)
of an element perpendicular to the plane of deformation. Thus, when $C>0$, the element collapses
in the direction perpendicular to the plane of deformation.
 For the case with initial conditions shown in Fig.~\ref{fig11}(a), $J_0>0$, $C<0$ and $D>1$. These 
conditions imply a non-singular solution as discussed above. On the other hand, for initial
conditions satisfying $J_0<0$ and $C<0$, the results are shown in Fig.~\ref{fig11}(b). 
As expected, the solution in this case also is singularity free. 

The effect of rotation in avoiding singularity has been discussed in detail for the two
dimensional case before. In the case of three dimensional deformations also, rotation may
play a role in rendering the solution non-singular. Note that $C$ in (\ref{c3digt0}) is unaffected by
$\omega$. Thus, the necessary condition for rotation to make the solution non-singular
is that $C<0$. Now, for $J_0>0$, the solution is singular if $D<1$. Thus, if $\omega_0$ is
chosen such that either it makes $D>1$, or makes $J_0<0$, singularity is removed. For a set of
two initial conditions which differ only in the value of $\omega_0$, the results are shown 
in Fig.~\ref{fig12}. It may be noted that here the singularity is removed by increasing $\omega_0$
in Fig.~\ref{fig12}(b) to make $J_0<0$.\\
\noindent{\bf Case 2}: $K_{ij}=k\delta_{ij}$ and $\beta=0$ 
\begin{enumerate}
\item{For $J>0$: \\
The exact solutions for $\theta$, $\sigma_{11}$, $\sigma_{12}$, $\sigma_{22}$ and $\omega$ are given by
\begin{eqnarray}
\theta &=& \frac{2{\sqrt k} \, p \, \sin 2 ( D-{\sqrt k}\,  t)} {[(q + p \cos 2 (D- {\sqrt k \, t})]} + \sqrt {k}
\tan  (C-{\sqrt k}\, t) ,  \\
\{\sigma_{11}, \sigma_{22}\} &=& \frac {1}{2} \left [ \,\frac{\theta}{3} -  \sqrt {k}
\tan  (C-{\sqrt k}\, t) + \frac{\{\, E, \, -E\,\} }{ [(q + p \cos 2 (D- {\sqrt k \, t})]} \right],  \\
\{\, \sigma_{12}, \, \omega \, \} & =& \frac{\{\, F, \, G \, \}}{ [(q + p \cos 2 (D- {\sqrt k \, t})]}  
\end{eqnarray}
where $p$ and $q$ satisfy $p^2- q^2 = 4k$. The integration constants are given in terms of initial conditions as
\begin{equation}
C=   \tan^{-1} \left [ \frac{\theta_0}{3\sqrt {k}}-\left(\frac{\, (\sigma_{11})_0 + \, (\sigma_{22})_0 } {\sqrt {k}}\right)\right],
\end{equation}
In order to obtain other integration constants, we first determine $q$ as 
\begin{equation}
q = \frac{1}{2B_0} \left[ - 4 k + \frac {B_0^{2}}{4k}\, \,  [\, \theta_0 -  \sqrt k \tan (\sqrt k C)\, ]^2 + B_0^{2} \, \right]\,.
\end{equation}
where $B_0$ is defined as,
\begin{equation}
B_0 = \frac{4k}{\sqrt{[\, (\sigma_{11})_0 - (\sigma_{22})_0 \, ]^2 + 4 J_0}}\,.
\end{equation}
The other constants can then derived as,
\begin{equation}
D =  \frac{1}{2} \tan^{-1} \left[\frac{B_0(\sqrt k \tan C-\theta_0)}{2\sqrt k (B_0 -q)}\right]\,.
\end{equation}
Now, $E, \, F$ and $G$ may be written in terms of $C$ and $D$ as follows
\begin{equation}
E=   (q+p\cos2D)\, [\, (\sigma_{11})_0 - (\sigma_{22})_0\, ]\,,
\end{equation}
\begin{equation}
\{\, F, \, G \,\}= \{ \, (\sigma_{12})_0, \,  \omega_0 \, \} \, (q + p \cos 2D).\end{equation}
These integration constants satisfy the following condition, 
\begin{equation}
 E^2  + 4 J_{0} \, [q + p \cos2D]^2 -16 k^2=0.
\end{equation} }
\item{For $J=0$ \\
\begin{equation}
\theta =  \sqrt {k}\,  [\, 
\tan  (C-{\sqrt k}\, t) + \tan  (D-{\sqrt k}\, t) + \tan  (E-{\sqrt k}\, t) ]\,,\end{equation}
\begin{equation}
\sigma_{11} = \left [ -\frac{\theta}{3} +  \sqrt {k}
\tan  (D-{\sqrt k}\, t)\right ],
\end{equation}
\begin{equation}
\sigma_{22} = \, \left [ \,-\frac{\theta}{3} +  \sqrt {k}
\tan  (E-{\sqrt k}\, t) \right ],
\end{equation}
\begin{equation}
\{\, \sigma_{12}, \, \omega \, \}  = \frac{\{\, F,\, G \,\}}{ \cos  (D- {\sqrt k \, t})\, \cos  (E- {\sqrt k \, t}) }
\end{equation}
The integration constant $C$ is same as for the case $J>0$ while others are given as,
\begin{equation}
D=   \tan^{-1} \left [ \, \frac{\theta_0  +3 \,(\sigma_{11})_0} {3 \sqrt {k}} \, \right],\, \, \, \, \, E=   \tan^{-1} \left [\,  \frac{\theta_0  +3 \,(\sigma_{22})_0} {3 \sqrt {k}}\,  \right],
\end{equation}
\begin{equation}
\{\, F,\, G \, \}= \{\, (\sigma_{12})_0, \,  \omega_0\, \} \,\cos D \, \cos E .
\end{equation}}
\end{enumerate}
The analytical solutions for the present case for $J<0$ are also possible with different 
conditions. The solutions for the case when  
$[\, (\sigma_{11})_0 -(\sigma_{22})_0\, ]^2 > 4 \vert J_0 \vert $ are same 
as for  $J>0$. It may be easily checked that all solutions with $K_{ij}\ne 0$ have finite 
time singularity.

In the general three dimensional case, due to a large number of variables, it may not be always  
possible to obtain analytical solutions of the kinematical quantities. However, one can  
obtain the numerical solutions for such cases. 

\section{Summary of results}

In this article, our primary purpose has been to find exact, analytical
solutions of the expansion, rotation and shear by solving the system of
equations involving these variables. We have also shown, in the cases of 
two and three dimensional media, how our solutions differ when the
initial conditions on the variables change.

A crucial result in both two and three dimensions, which, we believe, is
a generic characteristic of the system of equations is the existence of
a critical value for the initial condition on the rotation. Above and
below this critical value we have observed the presence and absence of
a finite time singularity in the solutions. However, it is also
observed that in three dimensions a change in the initial rotation alone
does not necessarily avoid a singularity. Additional necessary conditions
are also required to hold good in order to avoid a singularity. 

Characteristic  effects related to the presence of damping and
elastic forces are also evident in the behaviour of the ESR. 
For instance, in the presence of damping alone the area of a typical
element is found to shrink in time whereas for a purely elastic medium 
(without damping) the area is conserved over a period for oscillatory solution.
To help the reader appreciate our results we have  given a visual 
representation of our results by looking at the deformations of a 
two dimensional square element in time with typical initial conditions
above and below the critical initial rotation.    

A notable fact about our analysis in two dimensions is the identification
of specific regions in the space of initial conditions which give rise to 
either singular or nonsingular solutions of the ESR. We have not been
able to do so in three dimensions because of a larger number of variables
involved and also due to the absence of a unique variable `I' (as in two
dimensions). It will definitely be more illuminating if we could devise a way
of understanding the initial condition space in three dimensions in a manner
similar to the two dimensional case.

\section{Directions for future work}

Even though we have been able to solve the case of deformations in
two dimensional media exactly, the more interesting and realistic
case of three dimensional media in its full glory (i.e with all the
nine variables) remains an unsolved problem. It is surely a daunting
task analytically, considering the number of variables and the 
coupled and nonlinear nature of the equations. However, we hope to
be able to extract some useful information by numerically solving these
equations and making use of the analytical solutions described here
as checks on our numerics. 

A generalisation of the system of equations derived and solved in this
article concerns the situation where the evolution of the deformation
is not necessarily described by a linear equation. In other words, the
equation for $\ddot{\xi}$ may, in general, contain nonlinear terms of the
form $\xi^i \xi^j$ (quadratic), $\xi^i\xi^j \xi^k$ (cubic) etc. Such terms
would however lead to the presence of $\xi$ in the equations for
the ESR variables. A probable way to solve such a scenario is to include the 
equation for $\xi$ in the system of equations and solve for the whole
set either numerically or analytically. It is likely that an analytical
evaluation will not be possible unless we consider a largely simplified
model. 

We have mentioned in the Introduction that such equations could
describe a variety of mathematically similar situations.
The next obvious step in our study of such evolution equations for flows, is,
inevitably, the investigation of fluid flows in two and three dimensions.
Attempts towards analysing fluid flows are on and will be described in
future articles.

It may be apparent to the reader that the sequence of investigations being
carried out (i.e. for deformable media, fluid flow etc.) seem to
provide an indication towards the analysis of the full set of 
Raychaudhuri equations in a curved spacetime geometry for geodesic
and non--geodesic flows. That is a truly formidable task, which, as of now,
is not even a well--defined problem. As a preamble to such an exercise, it
may be worthwhile to look at a curved space generalisation of the 
deformation of two and three dimensional media and then fluid flows.

Finally, of course, we would like to see that our understanding of
the kinematics of deformable media as described in this article actually
does correspond to deformations in the real world. Our attempts at
visualisation through simulations might actually be realisable in simplistic
experiments. Note that one of the central results we have obtained
concerns the critical value of the initial condition on the rotation, on either
side of which we have nonsingular and singular solutions. This is also
a feature which was noted in a cosmological context, i.e., rotation
can prevent singularity formation in cosmology, though, spinning models
(like the Godel universe) had other problems and, moreover, they did not
correspond to a realistic cosmology. It would be instructive, if, in some way, 
we could show, even in a
toy experiment, the correlation between rotation and singularity formation.
\section*{Acknowledgments}
The authors thank Department of Science and Technology, Government of India for financial support through 
a sponsored project.

\noindent\section*{Appendix A} 
\begin{figure}[H]
\centerline{\includegraphics[scale=0.7]{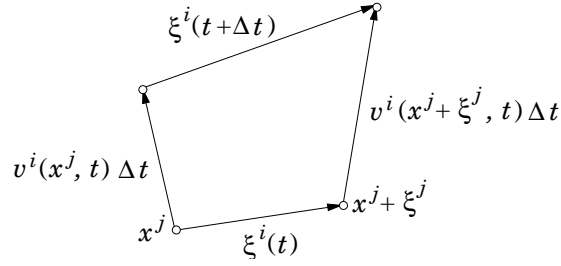}}
\caption{Schematic representation of evolution of deformation vector} 
\label{bijfig}
\end{figure}

Consider, at a time instant $t$, two infinitesimally separated points of a medium connected by the 
deformation vector $\xi^i$, as shown in Fig.~\ref{bijfig}. 
At a later time instant $t+\Delta t$, the
points move an infinitesimal distance along the respective instantaneous velocity directions, as shown. One can then write 
\begin{eqnarray}
&&\xi^i(t+\Delta t)=\xi^i(t)-v^i(x^j,t)\Delta t+v^i(x^j+\xi^j,t)\Delta t \nonumber \\
&\Rightarrow& \dot{\xi}^i=v^i_{\,\,;j}\xi^j  \nonumber \\
&\mbox{or}& \dot{\xi}^i=B^i_{\,\,;j}\xi^j \nonumber  
\end{eqnarray}
\noindent\section*{Appendix B} 
The equations (\ref{theta3})-(\ref{omega3}), for case 2 of three dimensional deformable medium having  $K_{ij}=k\delta_{ij}$ and $\beta=0$ with $\sigma_{13} =\sigma_{23}=0 $,  $ \omega_1 =\omega_2=0$, $\omega_3 =\omega$, reduce to the following set of equations, 
 \renewcommand{\theequation}{B-\arabic{equation}}
\begin{equation}
\dot \theta + z_1^{2} + z_2^{2}  + (\theta -z_1 -z_2)^2 + 2J + 3k  = 0, \,\, \, \, \label{b1} \end{equation}
\begin{equation}
\dot z_1 +  z_1^{2} + J + k = 0,  \label{b2}
\end{equation}
\begin{equation}
\dot z_2 +  z_2^{2} + J + k = 0,  \label{b3}
\end{equation}
\begin{equation}
\dot J +  2 (z_1 + z_2) J = 0.  \label{b4}
\end{equation}
where $ z_1 = \sigma_{11} + \frac {1}{3}{\theta}$ and  $z_2 = \sigma_{22} + \frac {1}{3}{\theta}$. The equation  (\ref{b4}) has the solution of the form $ J=J_0 e^{-2\int (z_1 +z_2)\,  dt}$. Further,  subtracting (\ref{b2}) and   (\ref{b3}) from    (\ref{b1}), one can calculate $z_1 + z_2$, and subtracting (\ref{b3}) from  (\ref{b2}) gives $z_1- z_2$, from which the different ESR variables can be calculated. In particular, for case 1 for three dimensional deformable medium, the equations (\ref{b1}) - (\ref{b3}) contain no term corresponding to the stiffness, and the solution for $J$ remains same.

\end{document}